# Alteration of Domain Microstructure by Ultrasound In Lithium Niobate.

I.V. Ostrovskii (1 and 2), N.O. Borovoy (2), O.A. Korotchenkov (2),

A.B. Nadtochii (2), R.G. Chupryna (2).

((1) Univ. of Mississippi, Oxford, MS, (2) Kiev Shevchenko Univ., Ukraine)

**Email: iostrov@phy.olemiss.edu**

**Abstract.**

A reorientation of the ferroelectric domains under an action of ultrasound in $LiNbO_3$ is observed for the first time. The involvement of the ferroelectric domain boundaries is experimentally identified by the analysis of X-ray reflection and crystal etching. The reorientation of the domains takes place under acoustic deformation of the order of $10^{-5}$ in megahertz frequency range.

## I. INTRODUCTION

In recent years, there has been much interest in the physical properties of lithium niobate owing to both the fundamental and applied research. [1-6] Its piezoelectric and optical properties are frequently used for applications, for example in the fields of acousto- and opto-electronics, fiber-optic telecommunications, information technologies, etc. As a basic material for active devices utilizing acousto-optic and piezoelectric properties, $LiNbO_3$ crystals frequently operate with a high mechanical/acoustical strain. However, a real crystal microstructure under dynamic stress so far has been given little attention. In particular the evolution of lithium niobate microstructure in acoustic field, including the ferroelectric domain walls and low-angle grain boundaries, has not been previously investigated. Such changes can result in a modification of the electromechanical quality factor Q.

This work aims to observe the changes in the crystal substructure, which are caused by the application of ultrasound; furthermore, the alteration of the physical properties of the samples caused by these changes is studied. We measured X-ray reflection, took the micrographs of the etching pits and figures, detected an acoustic emission from the samples under acoustic loading, and measured rf-admittance of the $LiNbO_3$ – plates for calculations of the electromechanical quality factor Q. From the X-Ray measurements, we take the rocking curves from the samples, which vibrate with gradually increasing amplitude. A double-crystal spectrometer was used for these measurements.

## II. EXPERIMENTAL TECHNIQUE.

Eight Y- and Z-cut $LiNbO_3$ rectangular plates are used in these experiments. The linear dimensions of the sample surfaces vary in the range 6 to 11 mm, and plate thickness ranges from 1 to 1.4 mm. The surfaces of the samples are metalized, and ultrasonic vibrations are excited by applying an rf voltage $U$ to this metallization. The fundamental shear vibrations propagating along the sample thickness are excited in the range from 1.6 to 2.4 MHz.

Ultrasound causes a mechanical strain inside the samples. Its amplitude $\varepsilon$ can be calculated by equation (1) [7] for a one-dimensional case, which is valid for fundamental vibrations in the plates with the thickness much smaller than surface dimension.

$$\varepsilon \approx U\sqrt{\frac{K^2C}{mv^2}\frac{4}{\pi}Q}\,, \qquad (1)$$

where *C* is a capacitance of the piezoelectric plate, *m* is a sample mass, *v* is a sound velocity, and *K* is the electromechanical coupling coefficient. The factor *Q* is experimentally determined from the measurements of rf-admittance Y versus frequency near fundamental vibrations at the resonance frequency $f_R$ . Admittance Y has a maximum at the $f_R$ and a minimum at so called anti-resonance frequency $f_A$ . Q = ($f_R$ / Δf), where Δf is a width of Y at 0.707 level of its maximum magnitude. The electromechanical coupling coefficient *K* can be experimentally determined via two frequencies – resonance $f_R$ and anti-resonance $f_A$ . As *K* increases, a larger difference ($f_A$ - $f_R$) is measured. All three parameters *Q, K* and *C* are experimentally measured at each applied rf-voltage *U*, which permits a plot of *Q* as a function of either the applied rf-voltage *U* or the acoustic deformation $\varepsilon$. From the experimental data taken with different crystals, we found the maximum strain amplitude to be about $\varepsilon \approx 2\times10^{-4}$.

To reveal the microstructure of the crystalline samples, microscopic examination of the optically polished surfaces of $LiNbO_3$ is made after chemical etching of the samples. The etching is done by a boiling solution of the acids $HNO_3$ and HF taken in a volume ratio of 2:1 at the temperature about 110°C, which has been shown to produce the best polishing and etching effect in $LiNbO_3$ [8]. A transmission optical microscopic examination makes it possible to test the crystals for extended defects after the etching is made.

We apply the X-ray diffraction taken from the sample surface at room temperature to estimate crystalline structure perfection. An X-ray tube producing the tungsten L-emission lines is used in this series of experiments. The data presented below are taken with the $L\alpha_1$ emission line dispersed by a plumbago (002) monochromator. The rocking curves are taken from the Y-cut $LiNbO_3$ plates. The sample is (1,1)-symmetrically oriented relative to the plumbago crystal. The diffracted X-ray photons are then counted by the electronics for a time interval of 25 s. We estimate an accuracy of the X-ray measurements to be about 0.2%, which is equivalent to 400 counts per second.

## III. RESULTS AND DISCUSSION

**3.1.** A typical rocking curve taken from the sample LNYR-B1 is shown in Fig. 1. It is known that the shape of a rocking curve is determined by the orientation and size of the coherent scattering blocks (CSB). An angle of CSB reorientation can then be estimated by implicitly suggesting that (i) the scattering curve from a single block is of a Gaussian shape, and (ii) all the CSBs have about the same linear size. In other words, the experimentally observed rocking curves can be approximated by a series of the Gaussian shapes with the same widths and peak intensities, which do not differ remarkably from each other. The result of such an approximation performed in the framework of the Levenberg-Marquardt algorithm is shown in Fig. 1. The experimental curve shape, presented by the points in Fig. 1, is well described by a series of four Gaussian shapes shown by the dotted lines. The solid line gives the sum of the Gaussians. Therefore, in this simplified picture, one may suggest that the observed rocking curve is formed by four CSBs reflecting the $L\alpha_1$-line beam.

The shape of the plot of Fig.1 is found to change quite remarkably after an ultrasonic excitation is applied to the crystal, as shown in Fig. 2 by plots $a$ and $b$. This effect is detected from different samples. In Fig. 2, the peak at $\theta = 0.105°$ increases for plot $b$. The samples, which demonstrate these effects, are classified as series I. Another group of samples, named series II, do not show significant variation in X-ray reflection under acoustic influence. For the series I, a decreasing of ultrasonic power shrinks the observed effect. It is important to note that some irreversibility is detected for the series I samples; in other words, *the rocking curve does not return to its original shape after the ultrasound is removed.*

In order to relate the experimental results to the scattering blocks, we need to discuss the likely diffraction of X-rays on strain gradients imposed by ultrasound. It is known that interference effects originating from the periodic potential of an ideal crystal strongly suppress the intensity of the diffracted rays. When applying a lattice strain, the Bragg-reflected intensity increases. As a consequence, the diffraction line becomes stronger and broader in the crystal subjected to longitudinal acoustic fields. [9] In contrast with the reported results, the data of Fig. 2 do not show such a behavior. The occurrence of a shear strain in the Y-cut $LiNbO_3$ can explain our result. Indeed, there should be no disturbance of the crystal lattice in the X-ray-tested subsurface layer leading to the lattice parameter gradients. Therefore, there should be no broadening of the reflection curve. Meanwhile, the shear ultrasonic vibrations would cause a strain in the crystal substructure deep in the sample. We note that the coherent scattering blocks of different types could in principle be reoriented, which in turn results in the evolution of the rocking curve shown in Fig. 2. This evolution is a strong indication that CSBs relax in acoustic field.

At this point, the corresponding approximation procedure may be used leading to the Gaussian peak positions marked $G_1$-$G_4$ and $G_1'$-$G_4'$ in Fig. 2 for the initial and acoustically loaded sample, respectively. By comparing the peaks, one can see that the rocking curve components shift towards smaller diffraction angles in the ultrasonically treated sample (Fig. 3-b). This shift is likely to be due to the thermal expansion of the lattice in the vibrating plate, since it becomes larger with acoustic power increase.

**3.2**. To determine the non-thermal effects of the ultrasonic loading, we have to analyze the relative positions of the Gaussian peaks instead of their absolute angular readings. Thus, the angular positions of the peaks $G_2$-$G_4$ are to be determined with respect to the position of $G_1$, and that of the peaks $G_2'$-$G_4'$ are to be determined with respect to the $G_1'$ peak. All possible combinations of the different pairs of Gaussians need to be taken into account. For any two Gaussian peaks, let us say ***i*** and ***j***, one can introduce the angle of reorientation $\beta_{ij}$, which occurs as a result of acoustically induced transformations. This angle can be given as:

$$\beta_{ij} = \left\| \left[ \Theta(G_i) - \Theta(G_j) \right] - \left[ \Theta(G_i') - \Theta(G_j') \right] \right\|. \qquad (2)$$

By applying this procedure to the data of Fig. 2, the average angle $<\beta_{ij}>$ is found to be 3.6" (that is 0.001°). It should be mentioned at this point that even much smaller angular shift of about 0.0003° is easily detected through the shape of the rocking curve. In addition, we made a computer simulation, which revealed that such a small change in a Gaussian angular position led to an obvious discrepancy between the experimental line shape and its Gaussian modeling.

**3.3.** The dependence of the $Q$ factor on increasing the vibration amplitude is remarkably different in the samples of series I and II. The $Q$ factor increase is rather large above $\varepsilon \approx 10^{-5}$ in the series I - curve 1 (sample LNZ-5). The enhancement in $Q$ may be as high as 50% (curve 1 in Fig. 3 at $\varepsilon \approx 4\times10^{-5}$). Further increase in acoustic amplitude causes the quality factor to drop (not shown in Fig. 3), which may be understood in terms of thermoelastic and dislocation losses.

The observed behavior is clearly contrasted with the behavior seen in the series II samples. Curve 2 in Fig. 3 (sample LNY-3) shows that the quality factor increases only slightly above $\varepsilon \approx 10^{-5}$, and then starts to decrease at $\varepsilon \approx 4\times10^{-5}$. Previously, it has been reported that the changes seen in curve 2 are indicative of the occurrence of dislocation-related losses.[10]

**3.4.** Analyzing the acid etched crystal surfaces may test the perfection of our samples. These are displayed in Fig. 4. The following defects are revealed: (1) dislocation etch pits presented by triangular shaped pits; (2) grain boundaries marked by arrows GB; (3) domain walls patterns shown by arrows DW.

It is evaluated, the samples of series II shows better quality compared to the ones of series I. The density of the dislocation pits is about $10^5$ $cm^{-2}$ in Fig.4 (series I sample), but it is much lower, about $10^3$ $cm^{-2}$ , for the samples of series II. Also the grain boundaries and the domain patterns are both undetectable in the series II samples.

Therefore, overall effect of ultrasound action on the acoustoelectric quality factor Q is higher in case of more defected series I samples (plot 1 in Fig. 3) in comparison with better quality samples of series II (plot 2 on Fig. 3). Under respectively high strain of $\varepsilon > 4\times10^{-5}$, the decrease in $Q$ takes place for both samples, which is due to acousto-dislocation interaction. Dislocation loops move in acoustic fields, and this is accompanied by increased acoustic losses.[11,10] The motion of dislocations in our samples can be directly observed by comparing the etching pits taken before (Fig. 4-a) and after (Fig. 4-b) acoustic loading. The newly developed dislocation pits are seen in the acoustically treated sample: next to the "New D" sign in Fig. 4-b.

Obviously, the rapid increase in the quality factor Q observed in series I samples (plot 1 in Fig. 3) is driven by another mechanism. As the series I displays a considerable amount of reoriented substructures (Fig. 2 and 4), the growth of $Q$ may be thought to be due to an alignment stimulated by ultrasound. Such an alignment would then decrease the acoustic loss, which is manifested by increased $Q$.

**3.5.** Since the samples used are of a pure single crystal phase, one may conclude that the differences in the ferroelectric domain structure of the series I and II samples govern the observed effect. In this framework, the presented data can be explained by acoustically stimulated motion of the domain walls interacting with point defects. There exists a piece of evidence, which supports this mechanism [3, 12-14]. The domain walls in as-grown samples are pinned by the defects with different strength. An applied acoustic load would unpin them and allow them to move to a new position determined by the surrounding defects with larger pinning strength. The domain reorientation caused by the acoustic field is confirmed by the etching experiments presented in Fig. 4-b and may explain the X-ray reflection experiments displayed in Fig. 2. Furthermore, due to the occurrence of strong pinning centers capturing the domain wall, one should observe partially irreversible changes of the rocking curve shape after the loading is terminated; and this is indeed observed.

With such an approach, the likely difference between the samples of the series I and II is that of their domain structure. We can therefore suggest that the extended domain structure seen in some of the series I samples (Fig. 4) would enhance the growth of $Q$. Appropriate growth is remarkably smaller in the samples with indistinguishable domain structure. Obviously, the other mechanism, which is likely to contribute to the growth of Q, is an acoustically stimulated reorientation of low-angle grain boundaries. The boundaries are present in the series I samples shown in Figs. 4, but they are seen to be absent in the series II samples. Therefore, the difference in the growth rate of $Q$ observed in curves 1 and 2 of Fig. 3 is indicative of substructure reorientation processes described above.

**3.6.** The presented approach is further supported by the estimation of the average size of scattering blocks determined by the X-ray reflection experiments (Fig. 2). Indeed, the average displacement of CSB may be approximated by

$$<x> = < l > < \beta_{ij} >, \quad (3)$$

where $< l >$ is the average size of CSB taken across the sample surface. By implicitly assuming $< x>$ to be roughly the plate vibration amplitude $u$, which can be estimated by the equation (4):

$$u = \varepsilon / k = \varepsilon \lambda / 2\pi = \varepsilon d / \pi, \quad (4)$$

where $k$ is the wave-number, $\lambda$ is the acoustic wavelength, and $d$ is the thickness of the plate, one obtains $<x> \approx 50$ Å at $\varepsilon = 5\times10^{-5}$. Taking $<\beta_{ij}> = 3.6''$ yields $< l >$ of the order of hundreds μm. This estimate may be considered to be in a good agreement with the previously reported size of the ferroelectric domains in $LiNbO_3$.[3]

## IV. CONCLUSIONS

We report a reorientation of ferroelectric domains in $LiNbO_3$ single crystals caused by a MHz-frequency ultrasound under the strain amplitude of the order of $10^{-5}$. The effect is independently confirmed by acoustically induced evolution of the X-ray reflection rocking curves, chemically etched crystal surfaces, and the acoustoelectric quality factor measurements.

The physical mechanism responsible for the interaction of the domains and ultrasound can be connected to the mechanical stress and piezoelectric field produced by a piezo-active acoustic vibration.

## Figure captions:

**Fig. 1**. The rocking curve of an as-grown $LiNbO_3$ plate (series 1 sample, LNYR-B1) without ultrasonic loading. Points – experiment, dotted line – approximating Gauss shapes, solid line – the sum of the shapes (see text for details).

**Fig. 2**. The evolution of a top part of the rocking curve shown in Fig.1: plot ***a*** – initial before ultrasound action, plot ***b*** – under ultrasonic loading with $\varepsilon \approx 5\times10^{-5}$ . Points – experiment, vertical lines marked $G_1$-$G_4$ and $G_1'$-$G_4'$ – the peak positions of the approximating Gauss shapes, solid lines – the sum of the shapes. Note the Gaussian lines $G_1$, $G_4$, $G_1'$ and $G_4'$ are responsible both for the top and wing parts of the rocking curve, that is why they appear to be slightly out of the upper narrow range of rocking angle.

**Fig. 3**. Acoustoelectric quality factor $Q$ vs acoustic strain $\varepsilon$ : plot 1 – series 1 sample LNZ-5, plot 2 – series 2 sample LNY-3.

**Fig. 4**. Acid etched cleavage surface in the series 1 sample (sample LNZ-5). The etching is performed before (**a**) and after (**b**) acoustic loading. **GB** and **DW** indicate grain boundary and domain wall, respectively. To the right of "**New D**" are shown the new dislocation pits, which are developed by acoustical treatment of the sample. Area "**New DW**" shows the location of a new domain wall, which occurs after ultrasonic action. The size of the sample regions is 165×90 $\mu m^2$.

**Fig. 1**

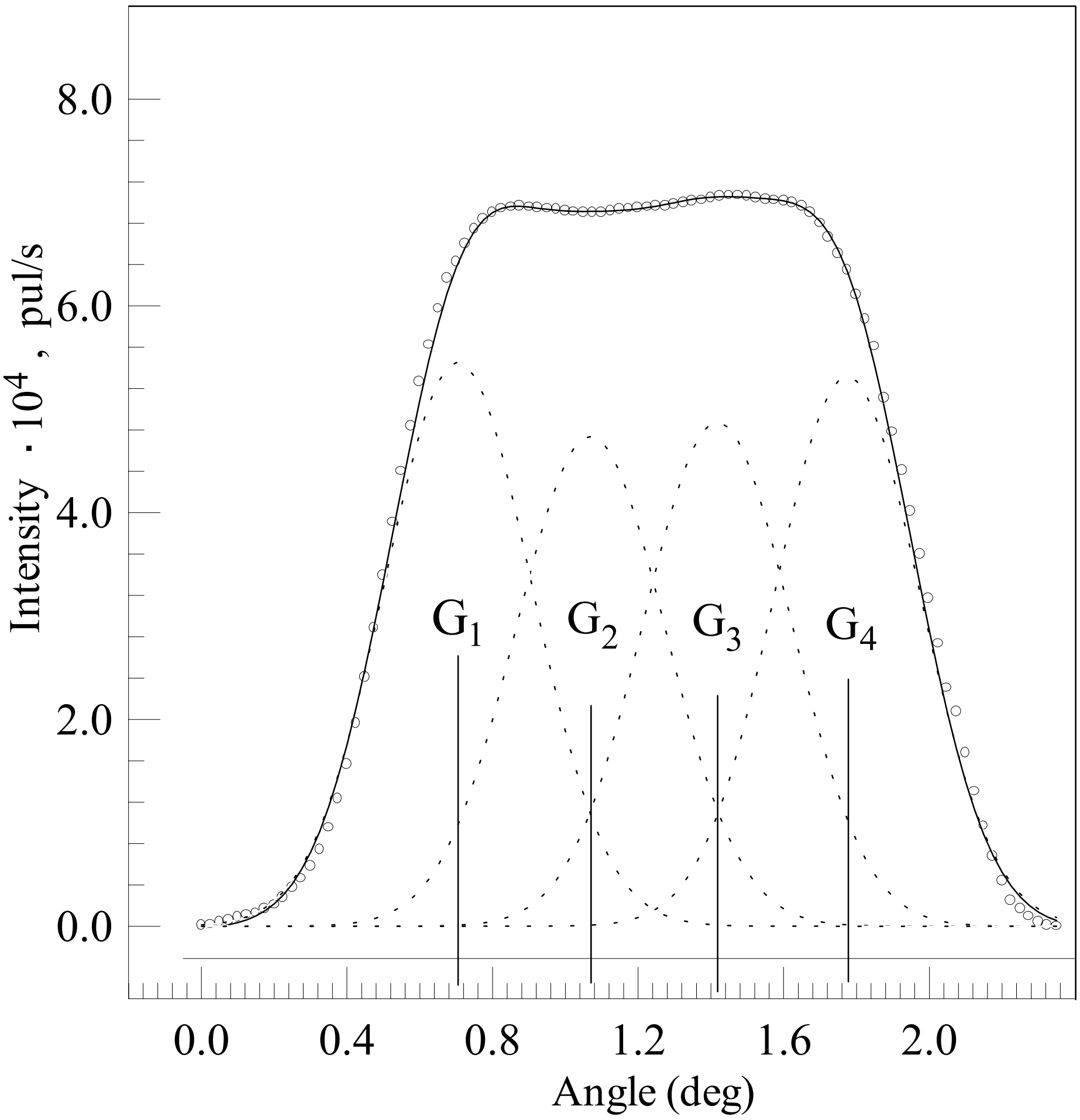

**Fig. 1**. The rocking curve of an as-grown $LiNbO_3$ plate (series 1 sample LNYR-B1) without ultrasonic loading. Points – experiment, dotted line – approximating Gauss shapes, solid line – the sum of the shapes (see text for details).

**Fig. 2**

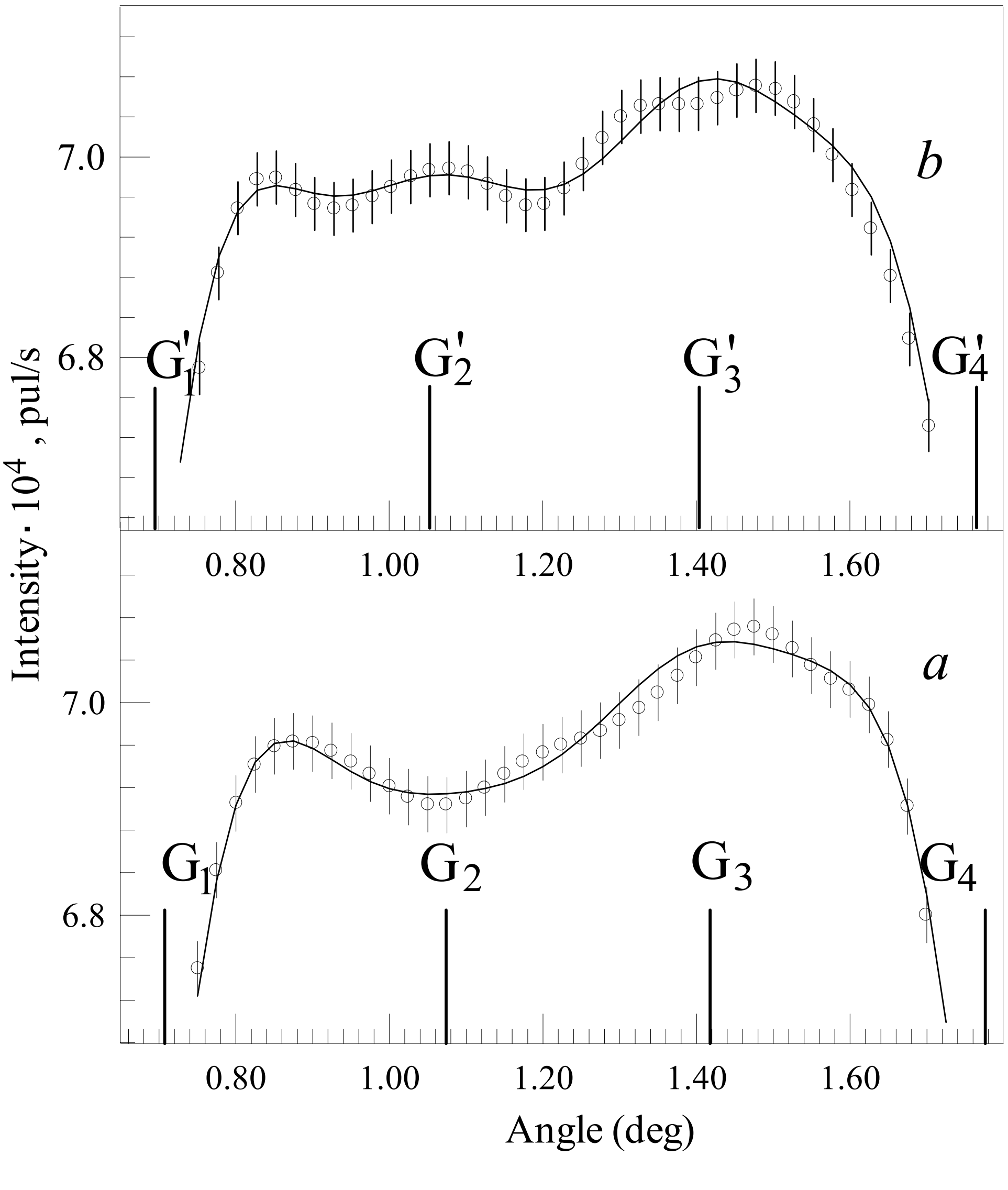

**Fig. 2**. The evolution of a top part of the rocking curve shown in Fig.1: plot ***a*** – initial before ultrasound action, plot ***b*** – under ultrasonic loading with $\varepsilon \approx 5\times10^{-5}$ . Points – experiment, vertical lines marked $G_1$-$G_4$ and $G_1'$-$G_4'$ – the peak positions of the approximating Gauss shapes, solid lines – the sum of the shapes. Note the Gaussian lines $G_1$ , $G_4$ , $G_1'$ and $G_4'$ are responsible both for the top and wing parts of the rocking curve, that is why they appear to be slightly out of the upper narrow range of rocking angle.

**Fig. 3**

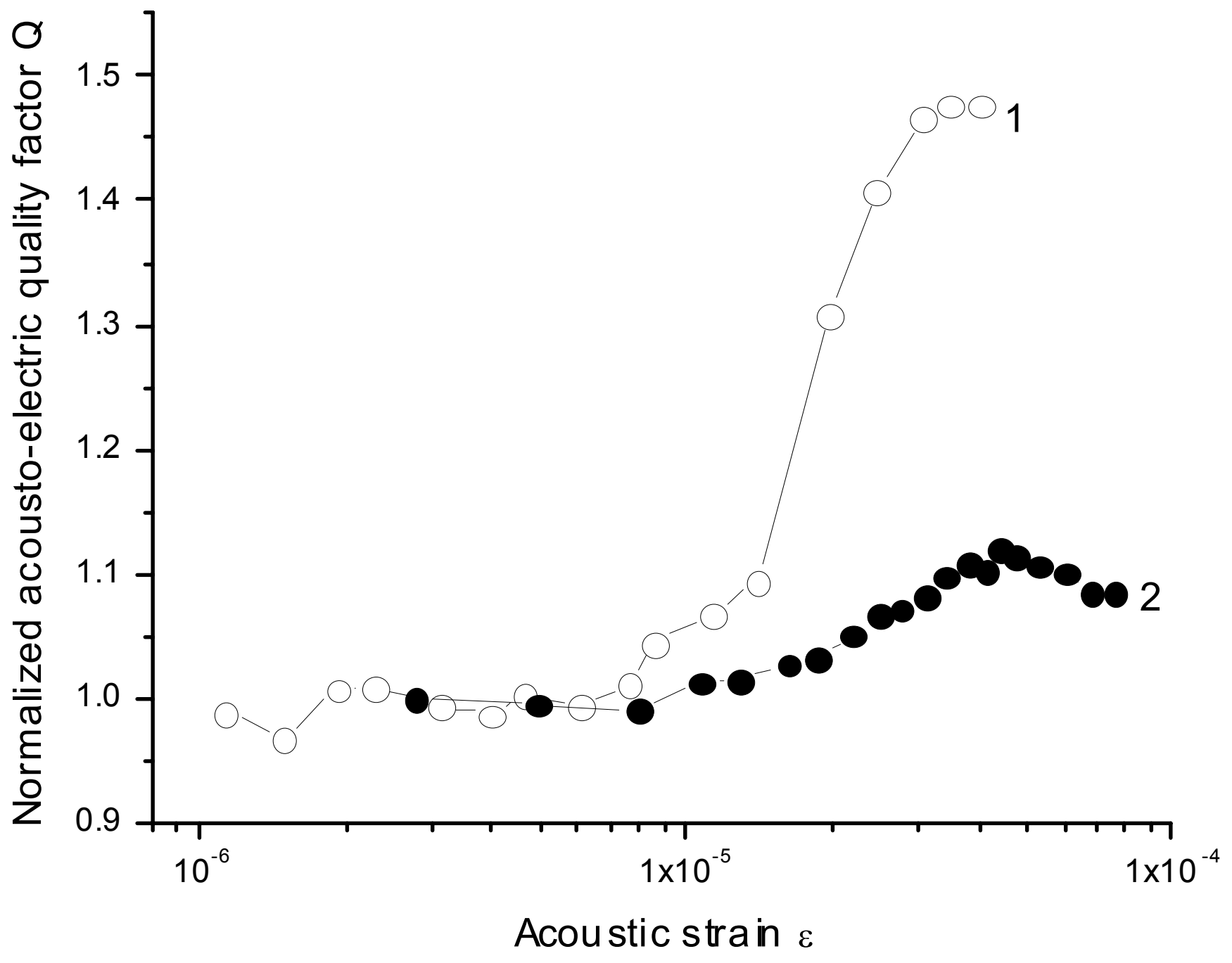

**Fig. 3**. Acoustoelectric quality factor $Q$ vs acoustic strain $\varepsilon$ : plot 1 – series I sample LNZ-5, plot 2 – series II sample LNY-3.

**Fig. 4-a.**

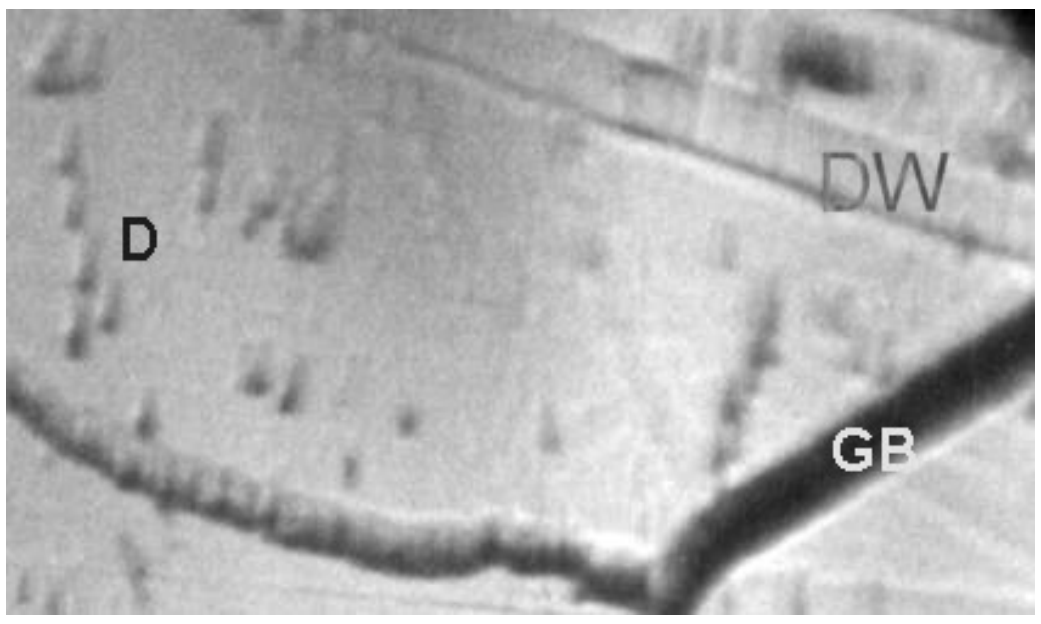

**Fig. 4-b.**

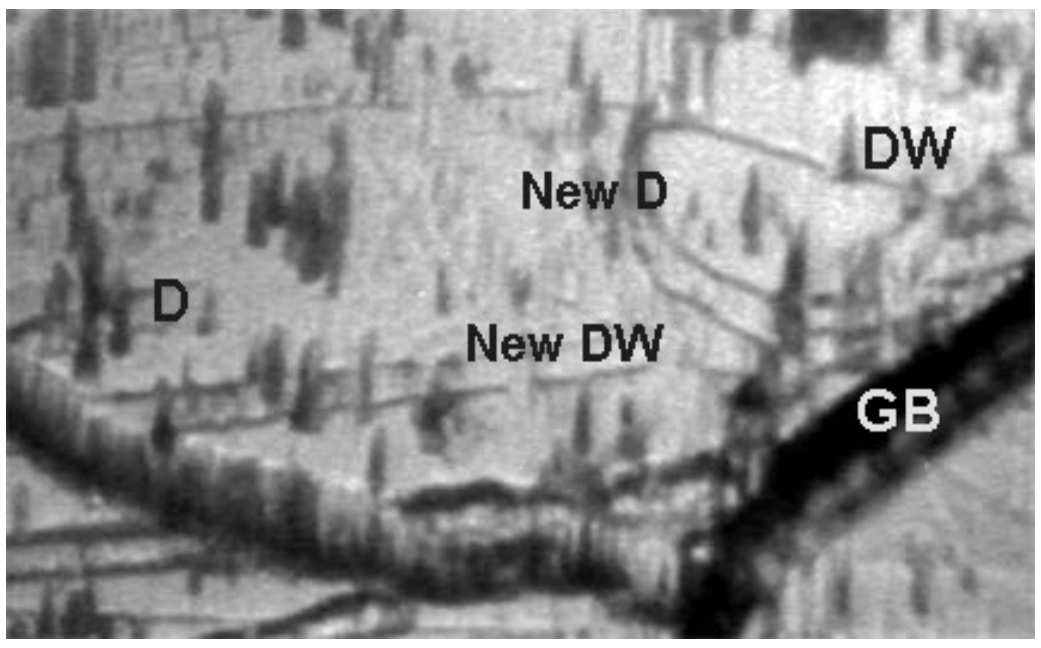